\documentclass[aps,pra,twocolumn,superscriptaddress,showpacs,floatfix]{revtex4-1}
\usepackage{lmodern,graphicx,amsmath,amssymb,amsfonts,multirow}
\usepackage[utf8]{inputenc}
\usepackage{gensymb}
\usepackage{longtable}

\usepackage{xcolor}

\begin{document}

\title{Continuous-wave virtual-state lasing from cold ytterbium atoms}

\author{Hannes Gothe}
\affiliation{Experimentalphysik, Universit\"at des Saarlandes, 66123 Saarbr\"ucken, Germany}
\author{Dmitriy Sholokhov}
\affiliation{Experimentalphysik, Universit\"at des Saarlandes, 66123 Saarbr\"ucken, Germany}
\author{Anna Breunig}
\affiliation{Experimentalphysik, Universit\"at des Saarlandes, 66123 Saarbr\"ucken, Germany}
\author{Martin Steinel}
\affiliation{Experimentalphysik, Universit\"at des Saarlandes, 66123 Saarbr\"ucken, Germany}
\author{J\"urgen Eschner}
\thanks{\href{mailto:juergen.eschner@physik.uni-saarland.de}{juergen.eschner@physik.uni-saarland.de}}
\affiliation{Experimentalphysik, Universit\"at des Saarlandes, 66123 Saarbr\"ucken, Germany}

\date{\today}

\begin{abstract}

While conventional lasers are based on gain media with three or four real levels, unconventional lasers including virtual levels and two-photon processes offer new opportunities. We study lasing that involves a two-photon process through a virtual lower level, which we realize in a cloud of cold ytterbium atoms that are magneto-optically trapped inside a cavity. 
We pump the atoms on the narrow $^1$S$_0$ $\to$ $^3$P$_1$ line and generate laser emission on the same transition.
Lasing is verified by a threshold behavior of output power vs.\ pump power and atom number, a flat $g^{(2)}$ correlation function above threshold, and the polarization properties of the output. 
In the proposed lasing mechanism the MOT beams create the virtual lower level of the lasing transition. 
The laser process runs continuously, needs no further repumping, and might be adapted to other atoms or transitions such as the ultra narrow $^1$S$_0$ $\to$ $^3$P$_0$ clock transition in ytterbium.

\end{abstract}

\maketitle

\section{Introduction}
\label{sec:Introduction}

The working principle of a laser relies on a pump process that creates population inversion between two atomic levels, and a cavity-enhanced decay between both levels. These levels are in general atomic energy eigenstates, but they may also be virtual levels that are created artificially by applying an additional field. Then, lasing happens as part of a multi-photon process that coherently combines the laser emission with the additional field.  

Several such processes are already known. A prominent example is lasing between dressed states of a two-level system, initially predicted by Mollow \cite{Mollow1972}; others are Raman lasing \cite{Grison1991, Tabosa1991, Hilico1992, Guerin2008, Vrijsen2011, Sawant2017} and four-wave mixing. It was shown in \cite{Guerin2008} that these three gain mechanisms can be realized in one and the same cold-atom system depending on the pump detuning and geometry. Another related class of gain mechanisms has been termed lasing without inversion \cite{Javan1957, Kitching1999, Mompart2000}, when the lasing action is not supported by population differences between the bare states, but through multi-photon processes and/or quantum interference. 

Such lasers extend the possibilities in designing and running laser sytems. They typically have low output powers, in the range of nW to $\mu$W, but they can be tuned by the pump light and can be spectrally very narrow if proper atomic transitions are used. Of particular interest are super-radiant lasers that work on narrow atomic transitions, to overcome limitations in cavity stability and achieve mHz linewidth \cite{Bohnet2012, Norica2016}.

Here we investigate a novel lasing mechanism that we observe in a cloud of cold, magneto-optically trapped Ytterbium atoms, and that relies on population inversion between an real upper level and a virtual lower level. Unlike Raman lasing, this mechanism occurs in a V-type configuration of atomic levels rather than a $\Lambda$-configuration. Unlike other laser processes observed on narrow transitions \cite{Bohnet2012, Norica2016}, our laser operates in continuous-wave mode without any repumping. 

This paper starts by introducing our experimental setup. Section \ref{sec:measured_data} verifies the laser properties of the cavity emission while section \ref{sec:raman_mechanism} explains the lasing mechanism. Finally, section \ref{sec:polarization} discusses the influence of the pump geometry and polarization on the output.

\section{Experimental setup}
\label{sec:setup}

\begin{figure}
\includegraphics[width=0.48\textwidth]{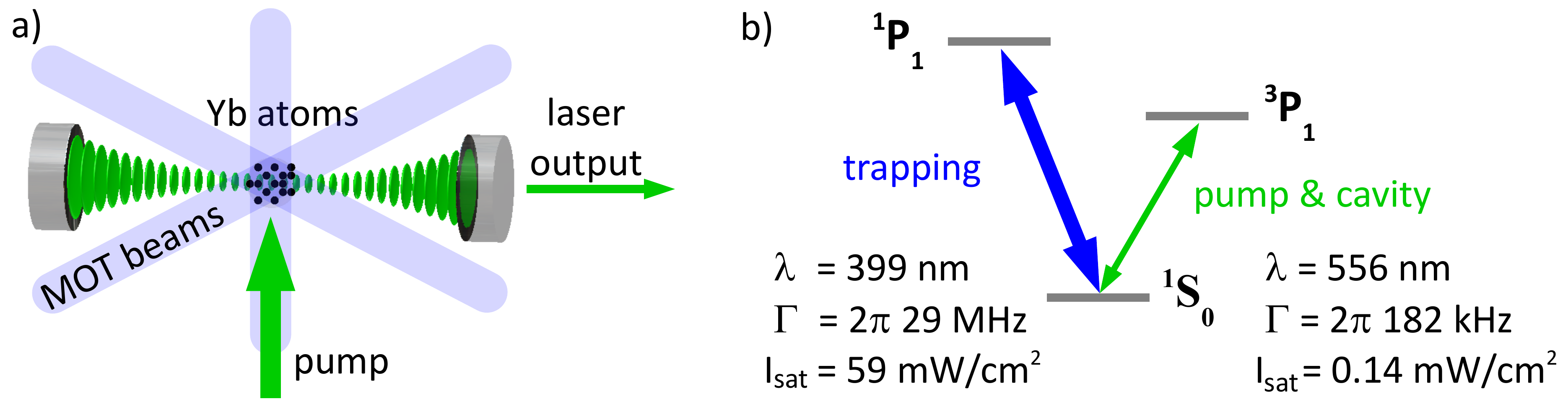}
\caption{(a) Schematic setup of the experiment. (b) Relevant transitions in $^{174}$Yb and their parameters. \label{fig:setup}}
\end{figure}

Our laser system is a cloud of $^{174}$Yb atoms that are trapped and laser-cooled inside a high finesse optical cavity. The atoms are first evaporated from an oven at 500$\degree$C, form an atomic beam that is decelerated by a Zeeman slower stage, and are finally captured in a magneto-optical trap (MOT). The trap operates on the $^1$S$_0$ ${\to}$ $^1$P$_1$ transition at 399\,nm (see Fig.\,\ref{fig:setup}) with a laser intensity of $0.5\times\mathrm{I_{sat}}$ per beam, a detuning around $-\Gamma$ and a magnetic field gradient of $36\,\mathrm{G/cm}$. We achieve a cloud temperature of 2\,mK \cite{Cristiani2010} and trap up to $10^7$ atoms at a cloud radius of $\sim1$\,mm.

The cavity consists of two high-reflectivity mirrors in Fabry-Perot configuration. It is resonant with the  $^1$S$_0$ ${\to}$ $^3$P$_1$ transition (see Fig.~\ref{fig:setup}b), has a linewidth of $\kappa=2\pi\times70\,\mathrm{kHz}$ (finesse $\mathcal{F}=55000$) and a waist radius of $w_0 = 90\,\mu$m. The cavity axis is tilted by 45$\degree$ with respect to the horizontal MOT beams. The mirror spacing of $4.78\,\mathrm{cm}$ allows us to operate the trap continuously inside the resonator and keep the atoms permanently overlapped with the cavity mode. The number of atoms is adjusted by a mechanical shutter that controls the flux from the atomic beam into the trap. Since the cloud diameter is about ten times larger than the cavity mode diameter, only $\lesssim$\,1\,\% of all trapped atoms, i.e. a few $10^4$ atoms, are overlapped with the cavity mode. The atom-cavity coupling rate is 2$\pi\times30$\,kHz corresponding to a coupling parameter of $C=0.1$ per atom. With our typical atom numbers we are in the collective strong coupling regime.

An additional laser beam at $\lambda=556$\,nm wavelength, with 7\,mW power and 2.4\,mm $1/e^2$-radius (corresponding to $\sim$280\,$\mathrm{I_{sat}}$ at resonance), whose frequency is controlled by a double-pass AOM stage, is overlapped with the vertical MOT beam and back-reflected at the top mirror, in order to pump the cloud on the  $^1$S$_0$ ${\to}$ $^3$P$_1$ transition. Under certain conditions, as detailed in the following sections, the atoms emit light into the cavity which subsequently leaks out through the mirrors. We analyse this light in several ways: with a photomultiplier tube (PMT, Hamamatsu H9656) we measure the output power, with a CCD camera we observe the transverse intensity distribution, and with two single-photon counting modules (LaserComponents COUNT-B50) we obtain the $g^{(2)}$-correlation function. Furthermore, for performing heterodyne measurements a detuned reference beam is overlapped with the cavity output, and the overlapped fields are measured on a fast photodiode. 

\section{Properties of cavity emission}
\label{sec:measured_data}

\begin{figure}
\includegraphics[width=0.48\textwidth]{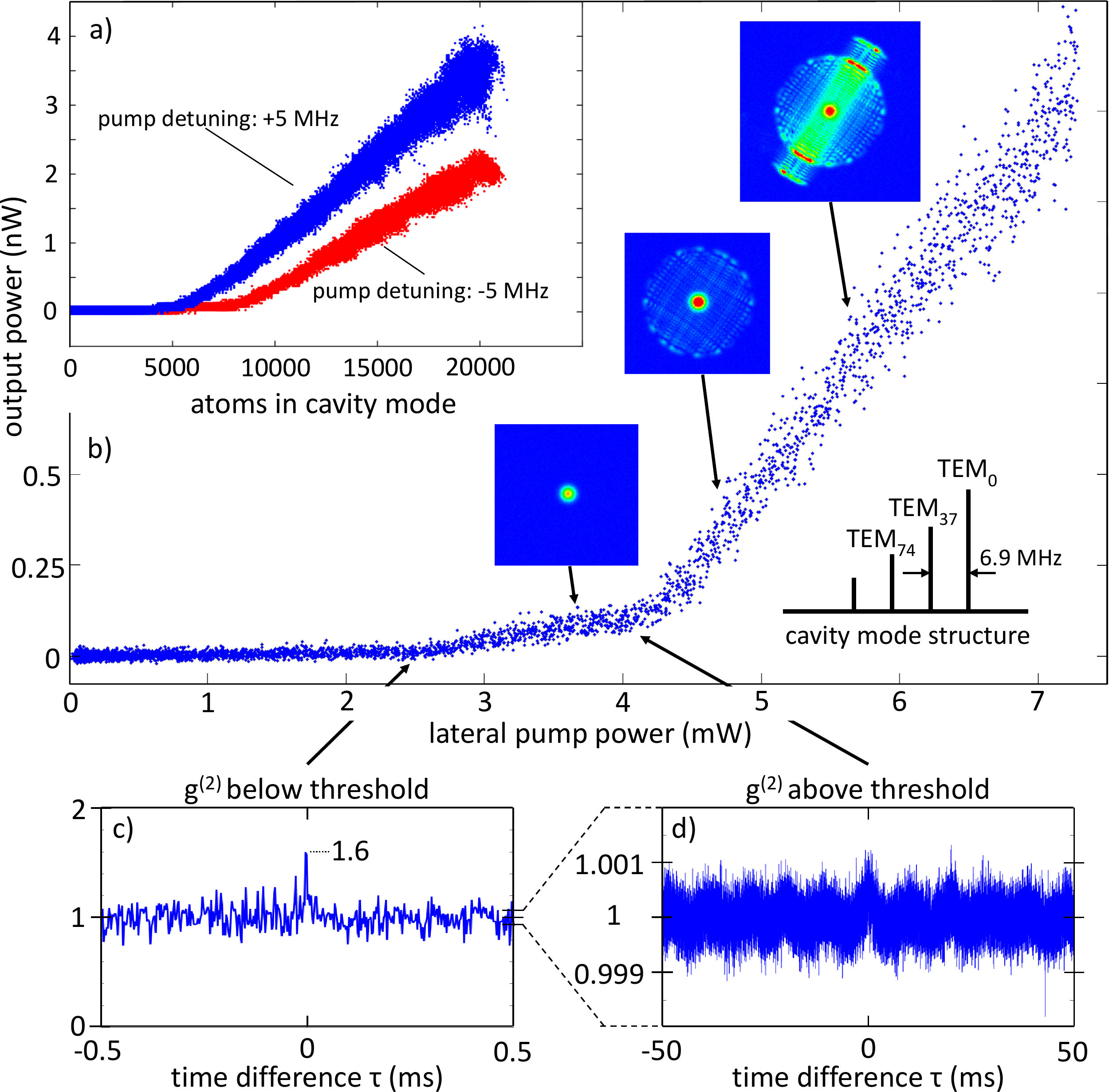}
\caption{
\label{fig:thresholds}
Cavity output power and $g^{(2)}$correlation function. a) Output power vs.\ atom number, exhibiting threshold behavior: at about 5000 atoms the fundamental mode starts emitting, for higher atom number the other modes emit as well. b)  Output power vs.\ pump power, exhibiting threshold behaviour. Single-mode output is found above 3\,mW, above 4\,mW light is also emitted in higher-order modes (TEM$_{37}$,TEM$_{74}$,~\dots). The cavity mode structure is depicted in the lower right corner. Threshold values and output power depend also on the pump detuning (see Fig.\,\ref{fig:detunings} and Sec.\,\ref{sec:polarization}). c) The $g^{(2)}$correlation function below threshold (2.5\,mW pump power, 270\,s, 500\,counts/s) shows a thermal peak of 1.6 at $\tau = 0$. d) The $g^{(2)}$correlation function of the TEM$_{0}$ lasing mode (4\,mW pump power, 22\,s measurement time, 500,000\,counts/s, 2.6\,$\mu$s bin size) exhibits a rise by less than 10$^{-3}$ at $\tau = 0$, plus some residual 50\,Hz noise.
}
\end{figure}

We observe bright emission from the cavity when the pump power and the atom number exceed a certain threshold (Fig.\,\ref{fig:thresholds}). Just above that threshold only the fundamental transverse cavity mode, TEM$_0$, is excited. For higher pump powers or atom numbers, several higher-order modes emit simultaneously. These modes are the spectrally closest ones to the fundamental mode, TEM$_{37}$, TEM$_{74}$ and TEM$_{111}$, with a mode spacing of 6.9\,MHz between each other. (We denote by TEM$_N$ the whole set of degenerate transverse modes TEM$_{n,m}$ for whom $n+m=N$.) 

The photon statistics in terms of the $g^{(2)}$ correlation function give further insight: when we choose a pump power and atom number such that stable output in only the fundamental mode TEM$_{0}$ is visible, we obtain a flat $g^{(2)}$ function (Fig.\,\ref{fig:thresholds}d). Clearly, the observed properties of the cavity light indicate lasing; 1\,nW of output power through one cavity mirror corresponds to about $6 \times 10^5$\,photons inside the cavity (taking into account that about 5\% of the total dissipated power is measured as laser output through one mirror while the rest is dissipated through other losses). 

\begin{figure}
\includegraphics[width=0.48\textwidth]{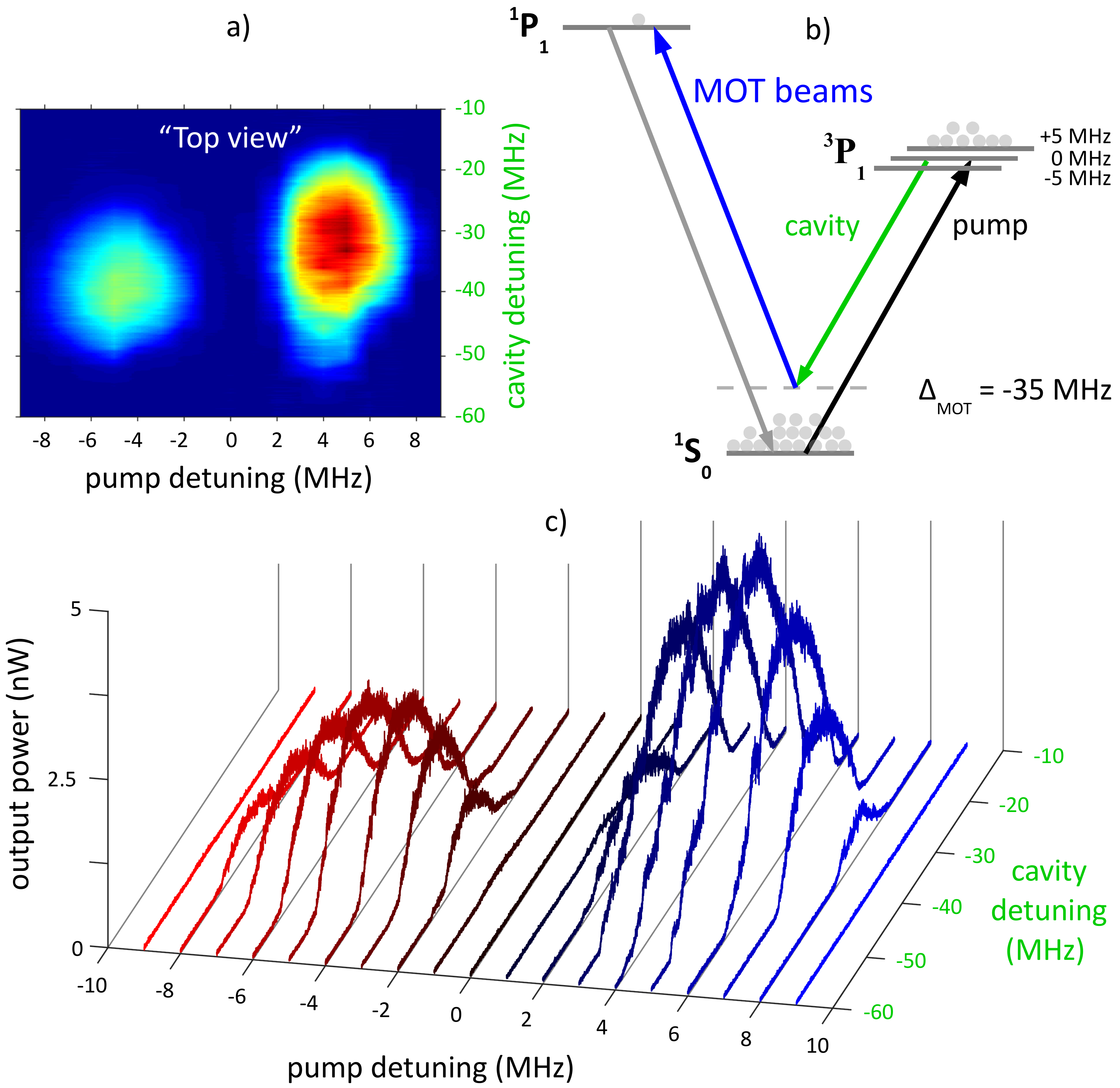}
\caption{Laser action vs.\ pump and cavity detuning. a) Overview of output power (color-coded) vs.\ detunings. The color scale ranges from dark blue (low power) to dark red (high power). b) Level scheme for the Raman lasing model, see sec.\,\ref{sec:raman_mechanism}. c) More detailed display of tha data in (a): the steady-state cavity output power is measured for various values of pump detuning (varied in steps of 1\,MHz by an AOM) and cavity detuning (scanned via a piezo actuator and calibrated before each scan by an on-axis reference beam with five known frequencies). All detunings are measured relative to the atomic transition. We observe two separated areas of emission, where the pump light is a few MHz red or blue detuned, and the cavity output is red-detuned by a few ten MHz. Lower output powers go along with single TEM$_{0}$ mode emission, while at higher power emission is multi-mode (see Fig.\,\ref{fig:thresholds}).  Note that this plot does not represent the emission spectrum as not all frequencies are emitted at the same time.}
\label{fig:detunings}
\end{figure}

Heterodyne measurement of the cavity output was also performed and yielded a linewidth of $\sim$240\,kHz, which is an unconclusive result: for a laser one would expect a value below the cavity linewidth of 70\,kHz. We explain this discrepancy by technical fluctuations: because of the low cavity output power, an integration time of several minutes is necessary, and slow frequency drifts between the cavity and the reference laser during this time enter into the heterodyne spectrum.

Important information is derived from the specific combinations of pump and cavity detunings, with respect to the atomic resonance, at which cavity emission occurs. We find lasing output in two regions of pump detuning (Fig.\,\ref{fig:detunings}a), one around $-5$\,MHz and another one around $+5$\,MHz. The corresponding cavity detunings are about $-30$\,MHz and $-40$\,MHz, respectively. In both cases the difference between input and output frequency totals $-35$\,MHz, which equals the detuning of the blue MOT beams. This leads us to the lasing mechanism presented in the next section.

\section{Lasing mechanism}
\label{sec:raman_mechanism}

The observation of the threshold behavior and the specific detuning dependence lead us to proposing a lasing mechanism involving a two-photon process through a virtual level. A level scheme with all relevant transitions, including the Zeeman sub-levels, is shown in Fig.\,\ref{fig:detunings}b. A fixed magnetic field is assumed in the figure; this holds for our experimental situation, as the pumped region was located slightly off the trap center, with an offset magnetic field of a few Gauss. 

The strong pump saturates the $^1$S$_0$ ${\leftrightarrow}$ $^3$P$_1$ (green) transition and creates significant population in the metastable $^3$P$_1$ levels (illustrated by gray dots). The red-detuned MOT laser provides not only cooling but also a virtual level above the $^1$S$_0$ ground state (dashed line), that is rapidly emptied. Lasing happens via this virtual level in a two-photon process whereby a green photon is emitted into the cavity and a blue MOT-beam photon is absorbed simultaneously. Spontaneous decay from $^1P_1$ back to the ground state closes the cycle. The process is most efficient when the pump detuning matches the Zeeman level shift, and when the combined detuning of pump and cavity matches the one of the MOT laser. This is consistent with the numbers found in the measurements: given $-35$\,MHz MOT detuning and assuming $\pm 5$\,MHz Zeeman shift, lasing emission happens around $\pm 5$\,MHz pump detuning and a cavity detuning of $-30$\,MHz, $-35$\,MHz and $-40$\,MHz. In the measurement of Fig.\,\ref{fig:detunings} the pump polarization did not allow the excitation of the $^3$P$_1$, $m=0$ Zeeman sublevel (see next section), therefore only the two outer peaks are observed.

The pump saturates the transition up to a detuning of $\sim$3\,MHz, which explains the observed pump width in Fig.\,\ref{fig:detunings}c. The much wider range of allowed cavity detunings corresponds to the broadening of the virtual level by its coupling to the 29\,MHz-broad $^1$P$_1$ level.

In the light of this explanation, the observed lasing process in ytterbium is clearly distinct from previously observed Raman lasing in cold atoms \cite{Hilico1992, Guerin2008, Vrijsen2011}, in that the two-photon process happens inversely, i.e.\ in a V-scheme between two excited states, with the lasing field providing the first amplitude and the applied (MOT) field the second. 

\begin{figure}[tbh]
\centering
\includegraphics[width=0.485\textwidth]{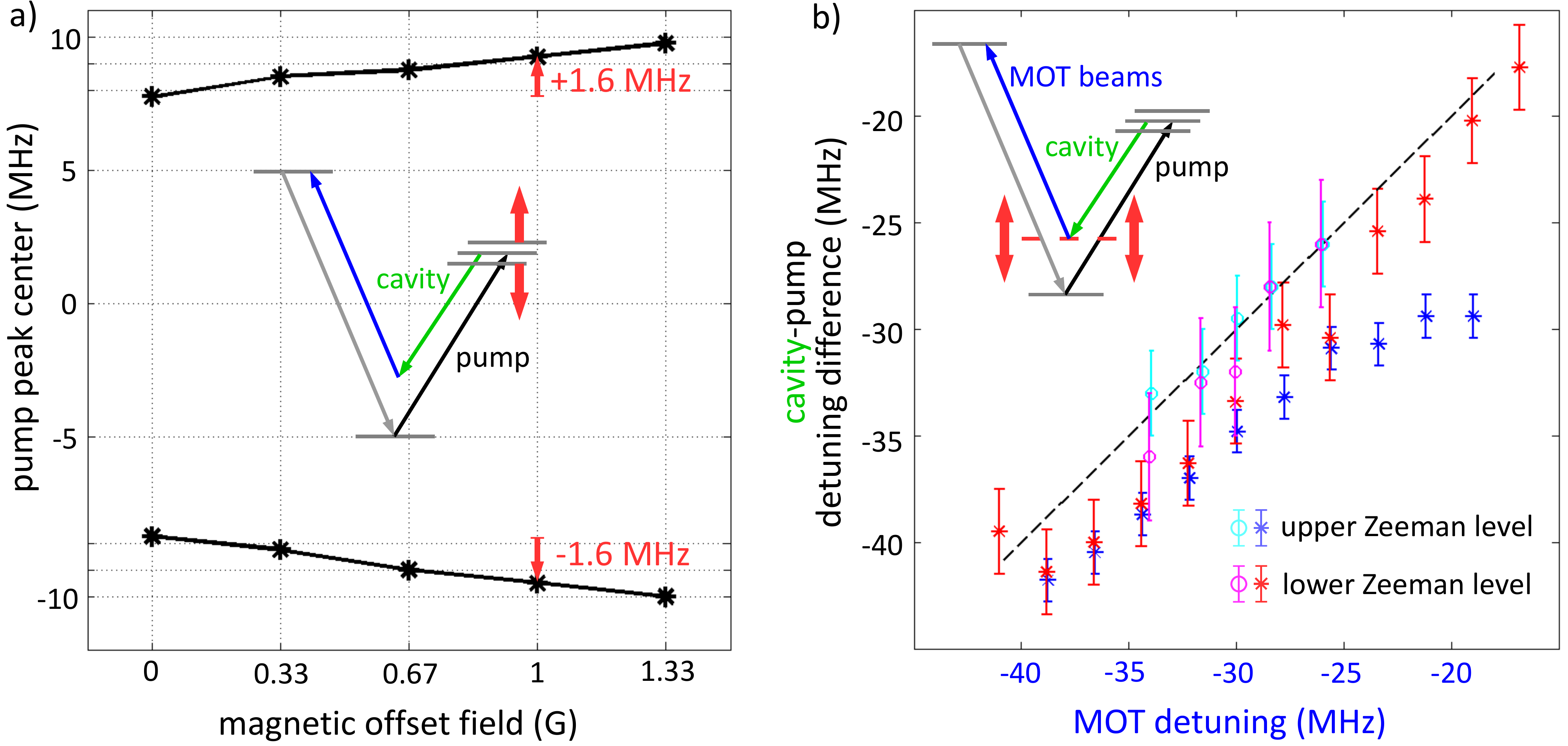}
\caption{
\label{fig:shifts} Effect on lasing of (a) a magnetic offset field and (b) the MOT detuning. For each set of parameters, a map like Fig.\,\ref{fig:detunings}a is recorded, and the values of optimum pump and cavity detuning are determined from the local maxima. The pump power is chosen such that only the TEM$_{0}$ mode emits. (a) Adding an offset magnetic field shifts the $^3P_1$ Zeeman sublevels (red arrows). The measured optimum pump detuning changes by 1.6\,MHz/G additional magnetic field. (b) The MOT detuning determines the virtual level (red broken line) of the two-photon process. Two data sets are shown (circles and stars) for MOT detunings between $-40$\,MHz and $-20$\,MHz; the optimum cavity detuning follows the MOT frequency. The observed offset is attributed to AC Stark shifts and to a locking point offset. The three blue outliers are caused by higher order mode emission that displaces the center of the overall emission peak.
}
\end{figure}

According to the suggested model, varying the MOT detuning should change the optimum cavity detuning by the same amount. Furthermore, an additional magnetic offset field should shift the $^3$P$_1$ Zeeman sublevels symmetrically and therefore modify the resonance condition for the pump light. Both assumptions are experimentally verified, as shown in Fig.\,\ref{fig:shifts}: a frequency shift of the MOT laser leads to an equal shift of the optimal cavity frequency (Fig.\,\ref{fig:shifts}b), and the optimum pump frequency shifts by about 1.6\,MHz per Gauss additional field (Fig.\,\ref{fig:shifts}a). The latter shift is slightly smaller than the one expected from the Land\'e factor of the $^3$P$_1$ level, $g=3/2$, which hints at additional Stark shift or cavity pulling effects.

\section{Pump geometry \& polarizations}
\label{sec:polarization}

In Fig.\,\ref{fig:thresholds} we see that pumping of the outer two $^3$P$_1$ Zemann substates ($\pm$5\,MHz) results in a cavity output, but pumping of the central state (0\,MHz) does not. When analyzing the polarization of the cavity output, we find that irregardless of the pump polarization the cavity emission is always circularly polarized: left-circular for a blue-detuned, and right-circular for red-detuned pump light. Both observations will be explained in this section by having a closer look at the pump geometry and the magnetic field. The magnetic field orientation and the dipole transition rules define which atomic transition ($\pi$ or $\sigma^{\pm}$) is excited by any specific input polarization and which polarization is emitted into the cavity. The magnetic field originating from our trap has quadrupole geometry: it vanishes at the trap center and increases (or decreases) linearly in all directions, see Fig.5a. Within the atomic cloud it varies accordingly in strength and orientation.

\begin{figure}[t]
\centering
\includegraphics[width=0.48\textwidth]{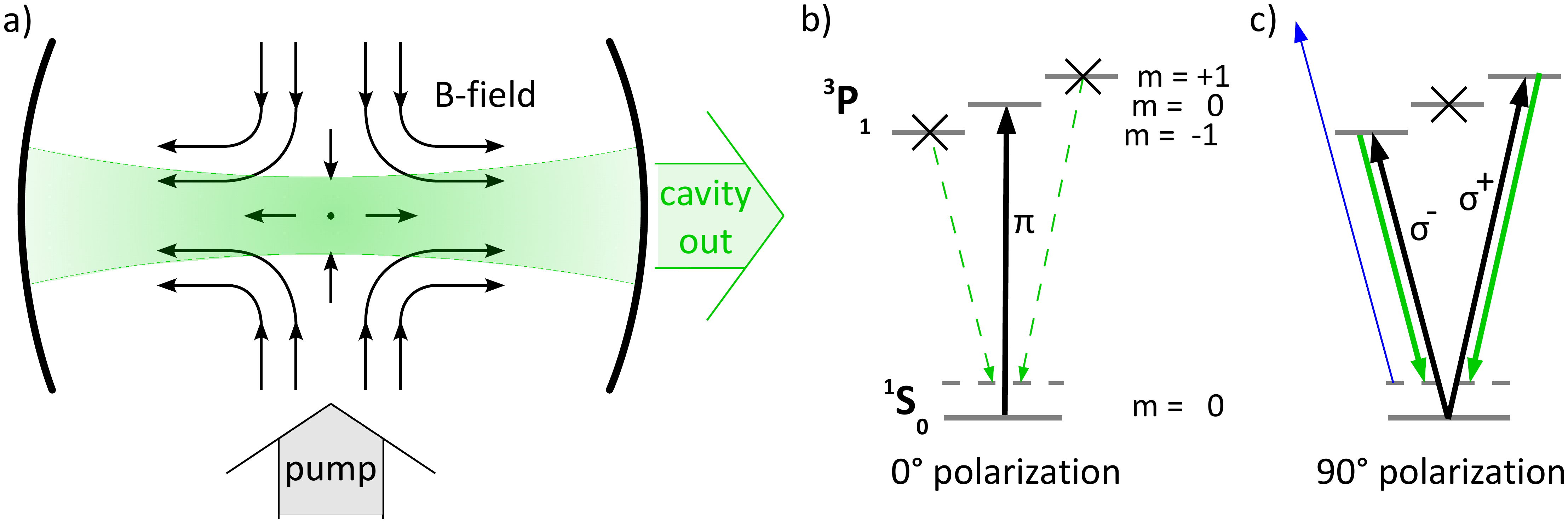}
\caption{\label{fig:polarization}
Geometry of trap and polarizations. 
(a) Cross section of the trapping area along the vertical and the cavity axis, showing the magnetic field of the MOT coils. Within the cavity mode volume, the field points predominantly along the cavity axis, thereby favouring $\sigma^+$- or $\sigma^-$-emission into the cavity, while $\pi$-emission is suppressed. 
(b) When the pump laser is linearly polarized along the cavity axis (0\degree) it excites the $\pi$-transition to the $^3$P$_1, m=0$ sublevel. The $m=\pm$1 sublevels are not populated, and no cavity emission occurs. 
(c) When the pump laser is linearly polarized perpendicular to the cavity axis (90\degree), it excites both $\sigma$-transitions to the $^3$P$_1, m=\pm1$ sublevels, making emission into the cavity mode possible. All other combinations of pump and output polarization are superpositions of these cases.}
\end{figure}

The relevant active lasing region is given by the overlap of the cavity mode, the atomic cloud and the pump field. Ideally this region is at the trap center with zero magnetic field, but in reality it may be shifted due to imperfections and misalignments leading to an offset in the magnetic field. As previously mentioned, this offset field and the pump polarization determine which atomic transitions are excited and which polarization is emitted into the cavity. A summary of the possible combinations is shown in Tab.\,\ref{tab:polarization}. All other settings are superpositions of these cases. The first two lines, where the magnetic field points along the cavity axis, are the most relevant ones for us, because this is the most likely outcome of our alignment procedure \footnote{For alignment we superimpose the cavity output and the MOT fluorescence on a CCD camera, looking along the cavity axis; this assures good alignment perpendicular to the cavity axis but not along the axis itself. We then move the pump beam until we see some cavity output. This may happen anywhere along the cavity mode resulting in an excited region that is likely to be at a magnetic offset field pointing along the cavity axis.}.

\begin{table}[t]
\begin{tabular}{c c ccccc ccccc}
\\
\multirow{2}{*}{B-field}\hphantom{m}& {pump}								& \multicolumn{5}{c}{excited}			 & \multicolumn{5}{c}{cavity output}\\
		  													& {pol.}								& \multicolumn{5}{c}{transitions}	 & \multicolumn{5}{c}{polarization}\\
\hline\hline	
\multirow{2}{*}{$\rightarrow$}	& 0\degree							&&---				&$\pi$& ---						&&&---&---&---&\\
																& 90\degree							&\hphantom{n}&$\sigma^-$&--- 	& $\sigma^+$		&\hphantom{n}&\hphantom{n}&R	&---&L&\hphantom{n}\\
\hline
$\uparrow$											& 0\degree\dots90\degree&&$\sigma^-$&---	& $\sigma^+$		&&&H 	&---&H&\\
	 
\hline
\multirow{2}{*}{$\bigodot$}			& 0\degree							&&$\sigma^-$&--- 	& $\sigma^+$		&&&V	&---&V&\\
																& 90\degree							&&--- 			&$\pi$& ---						&&&---&H	&---&\\
\hline
\end{tabular}
\caption{\label{tab:polarization} The combination of magnetic field orientation and pump polarization permits only specific $\pi$- 	or $\sigma^{\pm}$-transitions to populate the according Zeeman sublevels (m\,=\,0,\,$\pm$1) 
of $^3P_1$. Likewise, only specific polarizations can be emitted into the cavity mode to produce laser output.}
\end{table}

In this situation, according to Tab.\,\ref{tab:polarization}, we may only see right- and left-circular polarization or no output at all, depending on the input polarization. We verified these predictions by determining the laser threshold for various pump polarizations ($0\degree, \pm45\degree, 90\degree$ to cavity axis, and right- and left-circular). We find the lowest threshold, about 350\,$\mu$W, for 90\degree-excitation, an about two times higher value, 500-750\,$\mu$W, for $\pm45\degree$ or circularly polarized pump light, and no emission into the fundamental mode at $0\degree$. This agrees nicely with Tab.\,\ref{tab:polarization}. At a threshold of 2100\,$\mu$W, however, higher order modes start to emit even with $0\degree$ input; their larger mode volume covers also areas of different magnetic field orientations, and we assume this softens the previously discussed restrictions. 

The remaining configurations listed in Tab.\,\ref{tab:polarization} are obtained by changing the alignment between cavity mode, MOT and pump beam. By doing so we are, for instance, able to see emission from the otherwise suppressed central Zeeman transition that is listed in the last row of Tab.\,\ref{tab:polarization}. A complete polarization analysis will be addressed in future work.

\section{Conclusion}
\label{sec:conclusion}

We observe single- and multi-mode laser emission from a cold cloud of ytterbium inside a high-finesse cavity. The atoms are magneto-optically trapped on the broad $^1$S$_0\,-\,^1$P$_1$ transition and simultaneously pumped on the narrower $^1$S$_0\,-\,^3$P$_1$ intercombination line. This configuration allows for a V-type two-photon process between $^3$P$_1$ and $^1$S$_0$ with laser emission on the semi-forbidden intercombination line $^1$S$_0\,-\,^3$P$_1$. The observed process runs continously, needs no further repump lasers, and might equally be applied to the ultra-narrow $^1$S$_0\,-\,^3$P$_0$ clock transition.

Future work will be dedicated to linewidth measurements, in order to assess the potential for super-radiant laser applications. Other interesting aspects will be the dynamics of the multi-mode laser emission \cite{Wickenbrock2013, Horak2002}, as well as the interplay between the lasing action and the motional degrees of freedom of the trapped atoms, such as possible cooling or self-organization effects \cite{Xu2016, Jaeger2017, Domokos2002, Black2003}. 

\bibliography{lasing}

\end{document}